\documentclass[12pt]{article}
\newcommand{\be}{\begin{equation}}
\newcommand{\ee}{\end{equation}}
\newcommand{\beas}{\begin{eqnarray*}}
\newcommand{\eeas}{\end{eqnarray*}}
\newcommand{\bea}{\begin{eqnarray}}
\newcommand{\eea}{\end{eqnarray}}
\newcommand{\ba}{\begin{array}}
\newcommand{\ea}{\end{array}}
\newcommand{\nn}{\nonumber}

\newcommand{\al}{\alpha}

\newcommand{\de}{\delta}
\newcommand{\la}{\lambda}
\newcommand{\si}{\sigma} 
\newcommand{\n}{\nabla}

\begin{document}
\title{
{\bf Violating general covariance}
\author{Yu.\ F.\ Pirogov\footnote{e-mail: pirogov@ihep.ru}\\
{\it Theory Division, Institute for High Energy Physics,}\\ 
{\em Protvino, 142281 Moscow Region, Russia}}}
\date{}
\maketitle

\begin{abstract}
\noindent
The explicit violation of the general covariance on the whole and its
minimal  violation to the unimodular covariance specifically is
considered. The proper extension of General Relativity is shown to 
describe consistently the massive scalar graviton together with the
massless tensor one, as the parts of the metric. The bearing  of the
scalar graviton to the dark matter and  dark energy is indicated.
\end{abstract}

\section{Motivation}

The General Relativity (GR) is the viable theory of  gravity,
very robust in the underlying principles. It is known to consistently
describe  the massless tensor graviton as a part of the metric
field. This is insured by the general covariance (GC) which
serves as the gauge symmetry to eliminate the  degrees of freedom
contained in the metric in excess of  the massless tensor
graviton. Nevertheless, phenomenologically, the application of GR to
cosmology encounters a number of problems, superior of which are those
of the dark energy (DE) and the dark matter~(DM). In particular, to
solve the latter problem one adjusts usually the conventional or
hypothetical matter particles, remaining still in the realm of~GR.
The ultimate goal of DM being in essence to participate only in the
gravitational interactions, one can try to attribute to the aforesaid
purpose the additional degrees of
freedom contained in the metric, going thus
beyond~GC. With this in mind, I discuss in the given report the
self-consistent extension of GR, with the explicit violation of GC to
the residual unimodular covariance~(UC). In addition to the massless
tensor graviton, such an extension describes the massive scalar
graviton as a part of the metric field. The scalar graviton is
proposed as a resource of  the gravitational DM, as well as the scale
dependent part of DE.\footnote{The report is partly based on
ref.~\cite{Pirogov1}, where more details can be found.}

\section{GC and beyond}

\paragraph{Poincare group}

Let us first discuss the  problem of the GC violation from the point
of view of the
particle representation in the relativistic quantum mechanics. The
free particles are described by the irreducible finite-dimensional
unitary representations of the Poincare group
$ISO(1,3)$~\cite{Wigner}. The proper representations
$(m, s)$ are characterized by the mass  $m$ and spin~$s$. The massless
particles, $m=0$, possess the isotropic momentum 
$k_\mu$, $k\cdot k=0$.
The invariance group of the momentum (the ``little'' group)
proves to be $ISO(2)$, which is noncompact. The unitary
representations of the noncompact groups are known to be
infinite-dimensional, but for the scalar representations. Thus, for
a unitary representation of the Poincare group to
be finite-dimensional the  noncompact generators of the
little group (here the ``translations''of $ISO(2)$) should act
trivially on the representation. It follows thereof that the massless
particles of the spin
$s\ge 1$ should be  described not by the rays in a Hilbert space
but by the respective equivalence classes. This means that the theory
for the spin $s\ge 1$ should possess the invariance relative to
transformations within the proper equivalence classes, in other words,
be gauge invariant. Thus,  the gauge invariance is not a mere accident
but is in fact deeply rooted in the unitarity requirement for the
relativistic quantum theory.

Remind that the spin-one massless particle, say,  photon is described
by the transverse vector $\hat A_\mu(k)$, $k\cdot \hat A=0$. The
gauge transformations required for the triviality of the noncompact
generators, and thus for the unitarity,  is $\hat
A_\mu\to \hat A_\mu+\al k_\mu  $, with $\al(k)$ being a scalar. The
respective gauge group is~$U(1)$.
Due to this, one is left with the two-component photon
possessing  helicities $\la=\pm 1$.
Likewise, the spin-two massless  particle, the graviton, is described
by the transverse-traceless symmetric tensor $\hat
h_{\mu\nu}(k)$, with  $k^\mu \hat h_{\mu\nu}=0$ and 
$\hat h^\mu_\mu=0$~\cite{Vanderbij}.
The  gauge transformations required for the triviality of the $ISO(2)$
translations prove to be
\be\label{gt}
\hat h_{\mu\nu}\to \hat h_{\mu\nu}+ \xi_\mu k_\nu+\xi_\nu k_\mu,
\ee
with  $\xi_\mu(k)$ restricted by $k\cdot \xi=0$. The respective
three-parameter group corresponds precisely to UC.
Altogether, one arrives at the two-component graviton with the
helicities $\la=\pm 2$. 
Thus, UC is necessary and sufficient for the consistent description of
the massless tensor graviton. In this, the massive scalar
graviton can additionally be represented by the independent scalar
$\hat h(k)$ for the time-like momentum $k_\mu$, $k\cdot k=m^2> 0$.
The little group of the momentum being the compact  $SO(3)$, the
respective gauge transformations are trivial. 

One can abandon the reducibility requirement for the representation of
the massless tensor graviton, describing the latter at $k\cdot k=0$ by
the arbitrary transverse symmetric tensor $\hat h_{\mu\nu}(k)$, $\hat
h^\mu_\mu\neq 0$. For consistency, this requires the whole gauge
group, with arbitary $\xi_\mu$ corresponding to GC. Under these
transformations, the trace changes as $\hat h^\mu_\mu\to \hat
h^\mu_\mu+2k\cdot \xi$ and thus can be removed, leaving no scalar
graviton. It follows thereof that GC, with $\xi_\mu$ unrestricted,
though being commonly used and sufficient to consistently describe the
massless tensor graviton, is in fact redundant.

\paragraph{Field theory}

Let $x^\mu$, $\mu=0,\dots,3$, be the arbitrary observer's
coordinates. Let us now consider the same problem of the GC violation
in the framework of the Lorentz-invariant local field theory of the
symmetric tensor $h_{\mu\nu}(x)$. The latter is treated as a  part of
the dynamical metric field $g_{\mu\nu}(x)$. The effective field theory
of the metric  is to be built of the  metric itself  and its first 
derivatives $\partial_\la g_{\mu\nu}$ (as well as, generally, the
higher ones). Otherwise, one can use the
Christoffel connection $\Gamma^\la{}_{\mu\nu}(g_{\rho\si})$ which is
in the one-to-one correspondence with the first derivatives of the
metric. Now, $\Gamma^\la{}_{\mu\nu}$ is not a tensor and as such can
not generally be used as the Lagrangian  field variable. To remedy
this introduce the new field variable 
\be
\Omega^{\lambda}{}_{\mu\nu}= \Gamma^{\lambda}{}_{\mu\nu}-
\tilde\Gamma^{\lambda}{}_{\mu\nu},
\ee
with the compensating term 
$\tilde\Gamma^{\lambda}{}_{\mu\nu}$ being an external  nondynamical
affine connection. As the difference of the two connections,
$\Omega^{\lambda}{}_{\mu\nu}$ is the tensor and can thus serve as the
Lagrangian field variable. 
Generally, $\tilde\Gamma^{\lambda}{}_{\mu\nu}$ contains forty
components. Allowing for the
four-parameter coordinate freedom to bring four components
of $\tilde\Gamma^{\lambda}{}_{\mu\nu}$ to a
canonical form, there are still  left thirty six free
components. Thus, GC is completely violated. But for the field theory
of the metric to be consistent, at least the three-parameter residual
covariance is obligatory. This can be shown as follows.

Consider the linearized approximation  (LA) of the metric theory by
putting $g_{\mu\nu}=\eta_{\mu\nu} + h_{\mu\nu}$, 
with $h_{\mu\nu}$ being the
symmetric tensor field, $|h_{\mu\nu}|\ll 1$,  and $\eta_{\mu\nu}$
being the Minkowski symbol.
Specify some coordinates $x^\mu=(x^0,x^m)$, $m=1,2,3$,  and decompose
the symmetric Lorentz-tensor $h_{\mu\nu}(x)$ in terms of the $SO(3)$
fields as $h_{\mu\nu}=(h_{00}, h_{m0}, h_{mn})$. The second, namely,
the three-vector component in the decomposition possesses
the wrong norm, violating thus unitarity. The unitarity to be
preserved, the ``dangerous'' component should be eliminated. This
requires the three-parameter residual gauge symmetry, at the least.
In GR, one invokes the four-parameter gauge transformations 
\be
h_{\mu\nu}(x)\to  h_{\mu\nu}(x)+\partial_\mu \xi_\nu+\partial_\nu
\xi_\mu
\ee
with arbitrary $\xi_\mu(x)$ in accord with GC. Together with the
three wrong-norm components $h_{m0}$, these transformations eliminate
one more right-norm component. In the transverse gauge, $\partial^\mu
h_{\mu\nu}=0$, on the mass shell, $\partial\cdot\partial
h_{\mu\nu}=0$,  accounting for the
residual gauge freedom with the harmonic
parameters, $\partial \cdot \partial\xi_\mu=0$, one arrives
explicitly at  the two-component graviton. (Here one puts
$\partial\cdot \partial =\partial_\mu \partial^\mu$ and similarly for
any two vectors in what follows.) This procedure is quite reminiscent
of the electrodynamics where the vector field $A_\mu(x)=(A_0,A_m)$
possesses one, namely, scalar component with the
wrong norm. To eliminate this component the one-parameter gauge
symmetry $U(1)$ is required: $A_\mu\to A_\mu+\partial_\mu \al$, with
arbitrary $\al(x)$. In the transverse gauge, $\partial\cdot A=0$, on
the mass shell, $\partial \cdot \partial A_\mu=0$, with account for
the residual harmonic transformations, $\partial \cdot
\partial\al=0$, one is left explicitly with
the two-component photon.

To allow for some residual covariance one should 
reduce the number of the free components in
$\tilde\Gamma^{\lambda}{}_{\mu\nu}$. To this end, suppose that
$\tilde\Gamma^{\lambda}{}_{\mu\nu}$ is the
Christoffel connection for an external nondynamical metric
$\tilde g_{\mu\nu}$. The latter contains generally ten free
components. Allowing for the four-parameter coordinate freedom there
are left six independent nondynamical fields. Thus, the reduction of
the number of the fields is insufficient to
leave some residual covariance. The possible caveat is to confine
oneself to the contraction $\tilde\Gamma^{\la}{}_{\mu\la}$. 
Due to the relation 
$\tilde\Gamma^{\la}{}_{\mu\la}=\partial_\mu \sqrt{- \tilde g}$,  with
$\tilde g$ being the determinant of $\tilde g_{\mu\nu}$, the theory
depends in this case just on one nondynamical field.
The respective Lagrangian field variable becomes
\be
\Omega_\mu = 
\Gamma^{\lambda}{}_{\mu\lambda}-\tilde\Gamma^{\lambda}{}_{\mu\lambda}
=\partial_\mu\ln\sqrt{g/\tilde g}
\ee
In this marginal case, the nondynamical metric entering  only
through $\tilde g$, one can consider the latter just as a
scalar density of the proper weight. 
One can always choose the coordinates so that $\tilde g=-1$. Under the
variation of the coordinates $\de x^\mu= -\xi^\mu$, the scalar density 
$\tilde g$ varies as   $\de \sqrt {-\tilde g}=
\partial\cdot (\sqrt{-\tilde g}\xi)$.  
The residual covariance is that which leaves the canonical value
$\tilde g=-1$ invariant, requiring $\partial\cdot\xi=0$. This is 
the three-parameter UC. In this case, there is left one more
independent component in the dynamical metric. Precisely this extra
component corresponds to  the scalar graviton which can be
supplemented to the tensor graviton  not violating the consistency of
the theory. Note finally that the  dependence on the external
nondynamical field $\tilde g$ (more generally, on $\tilde
g_{\mu\nu}$) would tacitly imply that the metric Universe, contrary to
what is assumed in GR, should  be not a self-contained system and
could not entirely be described in the internal dynamical terms.

\section{Scalar graviton}

\paragraph{Lagrangian}

Let us study the  theory of the dynamical metric field $g_{\mu\nu}$
and the generic  matter field $\phi_{\rm m}$ with the generic action 
\be\label{GCV}
I=\int\Big(L_{\rm g}(g_{\mu\nu})+ \Delta L_{\rm g}(g_{\mu\nu},\chi)
+ L_{\rm m}( \phi_{\rm m}, g_{\mu\nu})+
\Delta L_{\rm m}( \phi_{\rm m}, g_{\mu\nu}, \chi)
\Big) \sqrt{-g}\,d^4x,
\ee
where 
\be
\chi=\ln\sqrt{g/\tilde g}.
\ee
Here $g=\mbox{\rm det\,}g_{\mu\nu}$ and  $\tilde g$ is a nondynamical
scalar density of the same  weight as $g$. Being the function of the
ratio of the two similar scalar densities, $\chi$  itself is the
scalar and thus can serve as the Lagrangian field variable.  
In the above,  $L_{\rm g}$ and  $\Delta L_{\rm g}$ are, respectively,
the generally covariant  and the GC violating contributions of the
gravity. Likewise, $L_{\rm m}$  and $\Delta L_{\rm m}$ are  the matter
Lagrangian, respectively, preserving and violating GC. All
the Lagrangians above are assumed to be the scalars.

Conventionally, take as~$L_{\rm g}$  the $\Lambda$-grafted
Einstein-Hilbert Lagrangian:
\be\label{EH}
L_{\rm g}=- \frac{1}{2}M_{\rm P}^2 \Big(
R(g_{\mu\nu})-2 \Lambda\Big),
\ee
where $R=g^{\mu\nu} R_{\mu\nu}$ is the Ricci scalar, with 
$R_{\mu\nu}$ being the Ricci curvature, and $\Lambda$ is the
cosmological constant. Also,
$M_{\rm P}=(8\pi G_{\rm N})^{-1/2}$ is the Planck mass, with 
$G_{\rm N}$ being the Newtonian constant. Present the scalar
graviton Lagrangian $\Delta L_{\rm g}$ as
\be\label{Ls}
\Delta L_{\rm g}= \Delta K_{\rm g}(\partial_\mu\chi, \chi)-\Delta
V_{\rm g}(\chi),
\ee
with $\Delta V_{\rm g}$ being the potential. 
In the   lowest order, the kinetic term $\Delta K_{\rm g}$ looks like
\be\label{K}
\Delta K_{\rm g}=\frac{1}{2} \kappa_0^2\,\partial
\chi\cdot\partial \chi ,
\ee
with $\kappa_0$ being a constant with the dimension of mass. 

The proposed extension of GR is more deeply rooted in the affine
Goldstone approach to gravity~\cite{Pirogov2}.
This approach is based on two symmetries: the
global affine symmetry (AS) and GC. AS terminates the theory in the
local tangent space, whereas GC insures the matching among the
various tangent spaces. Most generally, such a  theory depends
on an external nondynamical metric $\tilde g_{\mu\nu}$. This
dependence violates
GC and reveals the extra degrees of freedom contained in the dynamical
metric $g_{\mu\nu}$.  Call such
an extended metric theory of gravity the ``metagravity''.
Its minimal version, as considered in the report, depends  just   on 
$\tilde g$ and describes only the scalar graviton in addition to the
tensor one. Call  specifically the so reduced theory -- the
``scalar-tensor metagravity''.\footnote{This theory is not to be mixed
with the ``scalar-tensor gravity''~\cite{BD}. The latter is the
generally covariant extension of GR by means of a genuine scalar
field, which can not completely be absorbed by the metric. Also, the
theory proposed is to be distinguished from
the ``Unimodular Relativity'' based on UC but with the dynamical
metric  scale completely changed for the nondynamical one~\cite{Gal}.}
More generally, the metagravity 
can encompass also the vector graviton~\cite{Pirogov3}, though in this
case the unitarity is to be violated as well.

In the Lagrangian $\Delta L_{\rm g}$ above, $\Delta K_{\rm g}$
violates only GC, with $\Delta V_{\rm g}(\chi)$ violating also AS.
The GC violating part of  the matter Lagrangian, $\Delta L_{\rm m}$,
can be postulated in the simplest form as
\be
\Delta L_{\rm m}=- f_0J_{\rm m}(\phi_{\rm m}, g_{\mu\nu})\cdot
\partial \chi,
\ee
where $J_{{\rm m}\mu}$ is the 
matter current and $f_0$ is a scalar.  In the case when $f_0$ is a
constant, $\Delta L_{\rm m}$ above violates only GC, still preserving
AS. The possible dependence of $ f_0$ on
$\chi$  would reflect the  violation of AS, though 
still preserving UC. Allowing for $f_0\to 0$, independent of
$\kappa_0$, the matter sector can be
made as safe in confrontation between the theory and  experiment as
desired. For this reason, $\Delta L_{\rm m}$ will be disregarded in
what follows.

\paragraph{Classical equations}

By varying the action~(\ref{GCV}) with respect to $g^{\mu\nu}$,
$\tilde g$ being fixed, one  arrives at the modified gravity
equation:
\be\label{eomg}
G_{\mu\nu} = M_{\rm P}^{-2}\Big( T^{(\rm m)}_{\mu\nu} + \Delta
T^{(\rm g)}_{\mu\nu}\Big).
\ee
Here   
\be
G_{\mu\nu}=R_{\mu\nu}-\frac{1}{2}( R - 2\Lambda)g_{\mu\nu}
\ee
is the usual gravity tensor and   $T^{(\rm m)}_{\mu\nu}$ is the matter
energy-momentum tensor defined by~$L_{\rm m}$. The term $\Delta
T^{(\rm g)}_{\mu\nu}$ is the scalar graviton  contribution  looking
as follows: 
\bea\label{DT}
\Delta T^{(\rm g)}_{\mu\nu}&=&
\kappa_0^2  \Big(\partial_\mu\chi \partial_\nu\chi-
\frac{1}{2}\partial\chi\cdot \partial\chi g_{\mu\nu} \Big) 
+\Delta V_{\rm g} g_{\mu\nu}\nn\\
&&+\Big(\kappa_0^2\n\cdot\n \chi+
\frac{\partial \Delta V_{\rm g}}{\partial \chi}\Big) g_{\mu\nu}.
\eea
Mutatis mutandis, the first line of the equation above is the  
ordinary energy-momentum tensor of the scalar field.  The second
line  is the effective wave operator of the  field, with $\n_\mu$
being the covariant derivative, $\n_\mu \chi=\partial_\mu \chi$. This
line appeared solely due to the
dependence of $\chi$ on the metric and would be absent for the genuine
scalar field. We interpret the above contributions, respectively, as
those of the gravitational DM and the scale dependent part of DE,
caused by the scalar graviton. The latter having no
specific quantum numbers and undergoing only the gravitational
interactions, such an association is quite a natural~one.\footnote{The
above division on DM and DE is rather conventional. In particular in
the limit $\kappa_0\to 0$, the whole contribution of the scalar
graviton looks like DE.}$^{,}$\footnote{The other kinds of DM, if any,
are to be included in the matter Lagrangian.}

The r.h.s.\ of eq.~(\ref{eomg}) is thus proportional to the total
energy momentum of the nontensor-graviton origin, produced by the
nongravitational matter and  the scalar graviton.
Due to the Bianchi identity
\be
\nabla_\mu G^{\mu\nu}=0,
\ee
the total energy-momentum  is conserved: 
\be\label{cc}
\nabla_\mu (T_{\rm m}^{\mu\nu} +\Delta  T_{\rm g}^{\mu\nu} )=0,
\ee
whereas  the energy-momentum of the nongravitational matter alone,
$T^{(\rm m)}_{\mu\nu}$, ceases to conserve. 

To really solve the gravity equations one should impose the four
coordinate fixing conditions. E.g., one can choose the  canonical 
coordinates where $\tilde g=-1$, supplemented by the three more
independent conditions on the dynamical metric $g_{\mu\nu}$. As a
result,  $g_{\mu\nu}$ contains
generally seven independent components. Having solved the equations in
the distinguished coordinates one can recover the solution in the
arbitrary observer's coordinates.  Confronting the latter
solution with experiment one could conceivably extract 
the sought~$\tilde g$.

\paragraph{Linearized approximation}

To facilitate the problem of finding $\tilde g$ one could rely on~LA.
Not knowing $\tilde g$, guess from some physical considerations the
background metric $\bar g_{\mu\nu}$.  
Decompose the dynamical metric in LA as follows
\bea\label{WFA}
g_{\mu\nu}&=& \bar g_{\mu\nu}+h_{\mu\nu},\nn\\
g^{\mu\nu}&=& \bar g^{\mu\nu}-h^{\mu\nu} + {\cal O}((h_{\mu\nu})^2),
\eea
with $\bar g^{\mu\nu}$ being the inverse background metric. For the 
consistency, it is to be supposed  that $\vert h_{\mu\nu}\vert\ll 1$.
The indices are raised and lowered with
$\bar g^{\mu\nu}$ and $\bar g_{\mu\nu}$, respectively, 
so that $h^{\mu\nu}=\bar
g^{\mu\lambda}\bar g^{\nu\rho}h_{\lambda\rho}$, etc. Then one gets  
\be\label{chi}
\chi=  (h_0+h)/2+{\cal O}(h^2),
\ee
where $h\equiv \bar g^{\mu\nu}h_{\mu\nu}$
and $ h_0=\ln (\bar  g/\tilde g)$. 
The latter term is a  scalar parameter-field, 
not bound  in general to be small.
Physically, it reflects the discrepancy between the
background scale $\sqrt {-\bar g}$, which is at our disposal,  and
the nondynamical scale~$\sqrt {-\tilde g}$, which is given a priori.

The GR  Lagrangian in LA becomes as follows 
\be
L_{\rm g}=
\frac{1}{8}M^2_{\rm P}\Big((\bar\nabla_\lambda  h_{\mu\nu})^2
-2(\bar\nabla^\lambda h_{\lambda\mu})^2+ 2\bar\nabla^\lambda
h_{\lambda\mu} \bar\nabla^\mu h 
-(\bar\nabla_\lambda h)^2\Big) +{\cal O}((h_{\mu\nu})^3),
\ee 
with $\bar\nabla_\mu$ being the background
covariant derivative and $\bar\nabla_\mu h=\partial_\mu h$. 
The $\Lambda$-term is omitted here and in what follows.
For the respective gravity tensor, one gets
\be
G_{\mu\nu}=-\frac{1}{2}\Big(\bar\nabla\cdot\bar\nabla  h_{\mu\nu}
-\bar\nabla_\mu\bar\nabla^\lambda  h_{\lambda\nu}-
\bar\nabla_\nu\bar\nabla^\lambda  h_{\lambda\mu}
+\bar\nabla_\mu\bar\nabla_\nu h
\Big)
-\frac{1}{2}\Big(\bar\nabla^\lambda\bar\nabla^\rho
h_{\lambda\rho}
-\bar\nabla\cdot\bar\nabla h\Big)\bar g_{\mu\nu},
\ee
independent of $h_0$.  The Lagrangian above is invariant under the
gauge transformations
\be
h_{\mu\nu}(x)\to h_{\mu\nu}(x)+\bar\nabla_\mu \xi_\nu+ \bar\nabla_\nu
\xi_\mu,
\ee 
with  arbitrary $\xi_\mu$ corresponding to GC.
In particular,  one has $h(x)\to h(x)+2\bar\nabla\cdot \xi$. By this
token, $h$ can be removed, and thus $L_{\rm g}$, taken alone, 
does not produce any physical manifestations for the scalar graviton. 

The contribution of $\Delta L_{\rm g}$ to the gravity equations in
terms of $h_0$ and~$h$ can be read off
from eqs.~(\ref{DT}), (\ref{WFA}) and (\ref{chi}). 
This contribution is invariant only under the restricted gauge
transformations with $\bar\nabla \cdot\xi=0 $ or, otherwise, 
$\partial\cdot(\sqrt{-\bar g}\xi)=0 $. In the curved background, this
corresponds to the residual UC.
To solve the gravity equations one should impose on
$h_{\mu\nu}$ the three gauge fixing conditions, leaving thus seven
independent components. Comparing the solution with observations one
can conceivably extract thereof $h_0$ and, under the chosen $\bar g$,
the looked for $\tilde g$.

\paragraph{Quantization}

Assuming to have found $\tilde g$,  rescale the background metric
to adjust it to the external nondynamical scale, so
that $\bar g =\tilde g$.  Under this choice,
$h_0$ vanishes. The GC preserving part of the gravity Lagrangian stays
as before. The GC violating part reads 
\be
\Delta L_{\rm g}=\frac{1}{8}\Big(\kappa^2_0 (\bar\nabla_\lambda h)^2
 -\mu_0^4 h^2\Big)+{\cal O}(h^4),
\ee
with the potential supposed to be as follows
\be
\Delta V_{\rm g}(h)=\frac{1}{8} \mu_0^4 h^2+{\cal O}(h^4)
\ee
and  $\mu_0$ being a constant with the dimension of mass.
The Lagrangian $\Delta L_{\rm g}$ possesses only the
residual UC, with $\bar\nabla\cdot \xi=0$ insuring
$h\to h$. Normalized properly, the true field for the scalar graviton 
is $\kappa_0 h/2$, with the constant $\kappa_0$ characterizing thus
the scale of the wave function. At $\kappa_0\to 0$, the wave function
squeezes formally to dot. The other
free constant, $\mu_0$, characterizes the scalar
graviton mass, $m_0=\mu_0^2/\kappa_0$. 

Finally, the gauge fixing Lagrangian in the case of UC can be
chosen similar to ref.~\cite{Buchmuller}~as
\be
L_{\rm gf}=-\lambda(
\bar\nabla_\mu\bar\nabla^\lambda   h_{\lambda\nu}-
\bar\nabla_\nu\bar\nabla^\lambda  h_{\lambda\mu})^2,
\ee
with $\lambda$ being the indefinite Lagrange multiplier. This
condition
fixes three components in $ h_{\mu\nu}$, the scalar $h$ remaining
untouched. The forth independent gauge condition which is to 
be imposed in GR is now abandoned. It is superseded by
the GC violating term. The latter  looks superficially as the gauge
fixing term but with the definite coefficients. This is the principle
difference between the two kinds of terms. In the GC limit,
$\kappa_0\to 0$ and $\mu_0\to 0$, the given quantum theory becomes
underdetermined and requires one more gauge condition. For this
reason, the GC restoration is, generally, singular. 

Altogether,  one should study the present theory of the field
$h_{\mu\nu}$  in the curved background. As usually, this requires
the transition to the local inertial coordinates, what can in
principle be done. To facilitate the quantization procedure suppose
the Lorentzian background,
$\bar g_{\mu\nu}=\eta_{\mu\nu}$,  with the effect that 
$\bar \nabla_\mu= \partial_\mu$. The required ghost system is found
in this case in ref.~\cite{Buchmuller}. The respective propagator can
be shown to become
\be
D_{\mu\nu\rho\sigma}(x-x')=
\frac{1}{4}\Big(  P^{(2)}_{\mu\nu\rho\sigma}(\lambda)
\frac{1}{\partial\cdot\partial}+
\frac{1}{\epsilon_0^2}P^{(0)}_{\mu\nu\rho\sigma}
\frac{1}{\partial\cdot\partial+m_0^2} \Big)i\delta^4(x-x'),
\ee
where $
\epsilon_0=\kappa_0/M_{\rm P}$. 
The first term in the propagator corresponds to the massless tensor
graviton. The tensor projector $P^{(2)}_{\mu\nu\rho\sigma}$,
unspecified  here, corresponds to  the six
components of the  tensor graviton off the mass shell, as in GR.
The second term, with  the scalar projector
$P^{(0)}_{\mu\nu\rho\sigma}=\partial_\mu
\partial_\nu\partial_\rho\partial_\sigma/(\partial\cdot\partial)^2$,
describes additionally the scalar graviton. Altogether, the theory
describes  the seven propagating degrees of freedom 
reflecting ultimately the residual three-parameter UC.

In the limit $\kappa_0\to 0$, $\mu_0$ being fixed,
one gets for the scalar part of the propagator
\be
D^{(0)}_{\mu\nu\rho\sigma}(x-x')\simeq
\frac{1}{4\omega_0^2} P^{(0)}_{\mu\nu\rho\sigma}
i\delta^4(x-x'),
\ee
with $\omega_0\equiv\epsilon_0 m_0= \mu_0^2/M_{\rm P}$ being
finite. In this limit, the theory describes the massless tensor
graviton, as in GR, plus the contact scalar interactions. The
GC restoration limit, $\kappa_0\to 0$ and $\mu_0\to 0 $, is indefinite
in accord with the necessity of adding one more gauge
condition.\footnote{Conceivably, this is the particular manifestation
of a more general singularity at $\mu_0\to 0$ but $\kappa_0$ fixed,
corresponding to  the massless limit for the scalar graviton.}

\section{Conclusion}

In conclusion, the self-consistent extension of GR, with the explicit
violation of GC to the residual UC, is developed.  Being based on the
gauge principle, though with the reduced covariance, the extension is
as consistent theoretically as GR itself. In addition to the massless
tensor graviton, the respective theory -- the scalar-tensor
metagravity -- describes the massive
scalar graviton as the part of the metric field. The scalar
graviton is the natural challenger for the gravitational DM and/or the
scale dependent part of DE. The restoration of GR being
unattainable on the whole, the extension may be not quite safe
vs.\ observations. Its experimental consistency needs investigation.

I am grateful to Organizers for support and to W.~Buchm\"uller,
V.A.~Kuzmin,  V.A.~Ru\-bakov, and M.I.~Vysotsky for discussions.


\begin{thebibliography}{**}

\bibitem{Pirogov1} Yu.F.~Pirogov,  Phys.\ Atom.\ Nucl.\ {\bf 69}, 1338
(2006); gr-qc/0505031.
\bibitem{Wigner}
E.P.~Wigner, Ann.\ Math.\ {\bf 40}, 149 (1939). 
\bibitem{Vanderbij}
J.J.~van der Bij, H.~van Dam,  and Y.J.~Ng, Physica {\bf 116A}, 307
(1982).
\bibitem{Pirogov2} Yu.F.~Pirogov, 
Phys.\ Atom.\ Nucl.\ {\bf 68}, 1904 (2005); gr-qc/0405110.
\bibitem{BD} 
C.~Brans and R.H.~Dicke, Phys.\ Rev.\  {\bf 124}, 935 (1961).
\bibitem{Gal}  
D.R.~Finkelstein, A.A.~Galiautdinov, and J.E.~Baugh, J.\ Math.\ Phys.\
{\bf 42}, 340 (2001); gr-qc/0009099.
\bibitem{Pirogov3} Yu.F.~Pirogov, 
Phys.\ Atom.\ Nucl.\ {\bf 69}, 1622 (2006); gr-qc/0509093.
\bibitem{Buchmuller}
W.~Buchm\"uller and N.~Dragon, Nucl.\ Phys.\ {\bf B321}, 207 (1989).

\end{thebibliography}
\end{document}